\documentstyle[epsfig]{l-aa}

\def\ortho{\perp}

\def\d{{\rm d}}
\def\mA{{\cal A}}

\def\mD{{\cal D}}
\def\mS{{\cal S}}
\def\mV{{\cal V}}
\def\mV{{\cal V}}
\def\mH{{\cal H}}

\def\vk{{\bf k}}
\def\ii{{\rm i}}

\def\gam{{\bf \gamma}}
\def\mg{\big <}
\def\md{\big >}

\def\be{\begin{equation}}
\def\ee{\end{equation}}
\def\ba{\begin{eqnarray}}
\def\ea{\end{eqnarray}}
\def\beq#1{\begin{equation}#1\end{equation}}
\def\bea#1{\begin{eqnarray}#1\end{eqnarray}}

\input psfig.sty

\onecolumn

\begin{document}

   \thesaurus{2 (12.04.1; 12.07.1; 12.12.1)} 

 \title{The effects of source clustering on weak lensing statistics}

 \author{F. Bernardeau}

 \offprints{F. Bernardeau; fbernardeau@cea.fr}

 \institute{Service de Physique Th\'eorique, 
C.E. de Saclay, F-91191 Gif-sur-Yvette cedex, France}

\maketitle

\markboth{Source clustering effects}{F. Bernardeau}

\begin{abstract}
I investigate the effects of source clustering on the weak lensing
statistics, more particularly on the statistical properties of the
local convergence, $\kappa$, at large angular scales. The Perturbation
Theory approach shows that the variance is not affected by source
clustering at leading order but higher order  moments such as the
third and fourth moments can be.

I compute the magnitude of these effects in case of an Einstein-de
Sitter Universe for the angular top-hat filtered convergence.  
In these calculations the so-called Broadhurst and multiple lens 
coupling effects are neglected. The source clustering effect is
found to be particularly important when the redshift distribution is
broad enough so that remote background sources can be significantly
lensed by closer concentrations of galaxy sources.  The source
clustering effects are shown to remain negligible, for both the skewness
and the kurtosis, when the dispersion of the redshift of the sources
is less than about 0.15.
     

\end{abstract}

\section{Introduction}

The construction of gravitational distortion maps, for tracing
the large scale structure of the Universe, is a promising
tool for cosmology. For the first time, it would
indeed provides us with
an unbiased representation of the mass distribution in the universe
(Blandford et al. 1991, Miralda-Escud\'e 1991, Kaiser 1992).
And recent works tend to prove that indeed it should be possible
to have reliable distortion maps at the level expected
for the distortion induced by the large-scale structures
(Schneider et al. 1997a).

The cosmological interpretation of such maps is however
challenging. The main difficulty is that the background sources used
to make such measurements are very faint galaxies, whose  distances
and distribution is not necessarily well known.  It has been pointed
out in recent papers (Villumsen 1997,  Bernardeau et al. 1997, Jain \&
Seljak 1997) that it was crucial to know with a good accuracy the
redshift distribution of those sources. Their distances  determine
indeed the magnitude of the distortion effect: the more distant they
are the larger the effect is. The r.m.s. of the distortion is however
not the only information of cosmological interest to be extracted from
distortion maps. In particular the departure of the gravitational
convergence from a Gaussian statistics in case of Gaussian
initial conditions is an indicator of the amount
of nonlinearity reached by the cosmic density field (Bernardeau et
al. 1997, Jain \& Seljak 1997, Schneider et al. 1997b).  To get a
reliable description of such detailed analysis,  one has however to
take into account the possible effects produced by the intrinsic statistical
properties of the sources.

In this paper I investigate the effect of source clustering  for
high-order moment of the convergence. In Sect. 2,  the observational
schemes for the distortion and convergence fields are recalled, and
the couplings between the lens density fluctuations and the background
galaxy fluctuations are explicitly written.  The calculations are made
for an Einstein-de Sitter Universe only.  In Sect. 3, the
implications of such couplings for the skewness, third moment of the
convergence are investigated.  In Sect. 4, the expression of the
kurtosis, fourth moment, is computed, taking into account the source clustering
properties. The numerical applications, in Sect. 5,  are made for
different models of source distributions.

\section{Observational schemes for the distortion field}

\subsection{The filtering schemes}

In the weak lensing regime, and for small-size background objects,
the local gravitational distortion effects
can be entirely described by the local deformation matrix $\mA$.
This matrix expresses the local linear transform between 
the source and the image plane
induced by all the lenses present along a line-of-sight. 
Its inverse can  be related
to the second order derivative of the gravitational potential
through the equation,
\beq{
\mA^{-1}(\gam)={\rm Id}-\left\{
{3\over 2}\int_{0}^{\mD_s}\d\mD\,
{(\mD_s-\mD)\,\mD\over\mD_s}\,{\phi_{,ij}(\mD,\gam)\over a(\mD)}\right\}
\label{Ainv}
}
where ${\rm Id}$ is the identity matrix,
$\mD$ is the comoving angular distance along the line-of-sight, 
$a(\mD)$ is the expansion factor and $\phi_{,ij}(\mD,\gam)$
are second order derivatives of the local gravitational potential
at the position $(\mD,\gam)$ along the directions $i$ and $j$
orthogonal to the line-of-sight. Expression (\ref{Ainv}) is valid
for an Einstein-de Sitter only, but can be easily
extended to any background geometry. It is written for a given distance
of the source, $\mD_s$, that may vary for different
line-of-sights. 

In practice, the deformation matrix is not directly measurable.
The quantities that are directly accessible are the shape parameters
of the observed galaxies, $\mS^{I}$, in the image plane. For objects
with a large 
enough extension (compared to the width of the point spread function), 
this matrix can be related
to the shape matrix in the source plane, $\mS^{S}$, through,
\beq{
\mS^{I}=\mA^{-1}\cdot\mS^{S}\cdot\mA^{-1}\,{1\over \det(\mA^{-1})}.
}
Then the determination of the direction and amplitude of the
ellipticities of the galaxies gives a local estimation of the
deformation matrix\footnote{see however an alternative method
proposed by van Waerbeke et al. 1997, called the Pixel Autocorrelation 
Function, that does not use the shape matrices.}. 
Because of the intrinsic ellipticities
of the sources, the cosmic distortion can be detected only when 
a large number of galaxies are taken into account. Averaging over few
hundred of galaxies, distortion signals down to a few percents are
in principle detectable (Blandford et al. 1991).

The particular quantity that can thus be 
reconstructed is the local filtered convergence,
$\kappa$. It is directly related to the trace of $\mA^{-1}$ with,
\beq{
\kappa=1-{\rm tr}\left(\mA^{-1}\right)/2.
}
This quantity can be obtained directly from the observed galaxy
shapes, when it is filtered with a compensated filter (i.e. convolved
with a function of zero integral, see Kaiser 1995, Schneider et
al. 1997b).  In general, however, it is always possible to obtain a
convergence map from a distortion map by solving a differential
equation (Kaiser 1995).  In the following I will therefore focus my
analysis on the statistical properties of the local convergence,
filtered by a top-hat window funtion.

In the expression (\ref{Ainv}) the relation between the deformation matrix
and the gravitational potential is given for a unique
distance of the source plane. But actually the measured convergences
result from averages made over many background galaxies
that can be at different distances. More specifically
the measured local convergence at scale $\theta_0$ is 
obtained from background objects taken for instance
in a solid angle of radius $\theta_0$, so that it reads,
\beq{
\kappa_{\theta_0}={1\over N_s}\,\sum_{i=0}^{N_s}
\kappa(\gam_i),
\label{DisS}
}
where $\gam_i$ is the direction of the $i^{\rm th.}$ source galaxy.
The number density of source
galaxies that can be used is about 40 per arcmin$^2$
(with the usual deep exposures in the I band). For a filtering
radius of $20'$, we expect then to have about $50\,000$
galaxies. This number is large enough to assume that the discretization
of the background field is not important. We can then work in the
coutinuous limit in the source plane. The number density
of sources at distance $D_s$ and in the direction $\gam$
can be written,
\beq{
n_s(\mD_s,\gam)=n_s(\mD_s)\,(1+\delta_s(\mD_s,\gam)),
}
where $n_s(\mD_s)$ is the average number density
of sources\footnote{Note that for an Einstein-de Sitter Universe
the angular distance at $z\to\infty$ equals 2.}(that 
fulfill the selection criteria at a given distance), 
and $\delta_s(\mD_s,\gam)$ is their local density contrast.
The density $n_s$ is normalized to unity, $\int_0^2\d\mD_s\,n_s(\mD_s)=1$.
Writting the equation (\ref{DisS}) in the continuous limit we have,
\beq{
\kappa_{\theta_0}=-{3\over 2}\,{\int\d^2\gam\,W_{\theta_0}(\gam)\,
\int_{0}^{2}\d \mD_s\,\int_{0}^{\mD_{s}}
\d \mD\,{\mD\,(\mD_s-\mD)\over a\,\mD_s}\,
\delta_{\rm mass.}(\mD,\gam)\,n_s(\mD_s)\,(1+\delta_s(\mD_s,\gam))\over
\int\d^2\gam\,W_{\theta_0}(\gam)\,\int_0^2\d\mD_s\,n_s(\mD_s)
\left[1+\delta_s(\mD_s,\gam)\right]}.
\label{surf1}
}

In this expression, the two filters are a priori both top-hat filters,
\bea{
W_{\theta_0}(\gam)&=1\ \ \ {\rm for}\ \ \vert\gam\vert\le\theta_0;\nonumber\\
W_{\theta_0}(\gam)&=0\ \ \ {\rm for}\ \ \vert\gam\vert > \theta_0.
}

Note however that in general the filters are not necessary
the same.  For instance one can use a compensated
filter for the convergence, and still use a top-hat window 
for the selection of the sources. More generally, instead
of giving an equal weight to all sources  in a given area, it is
always possible to weight them proportionally to the inverse of
the local density. We would then have,
\beq{
\kappa_{\theta_0}^{\rm surf.}={\sum_{i=0}^{N_s} w_i\,\kappa(\gam_i)\over
\sum_{i=0}^{N_s} w_i},
\label{DisS2}
}
where $w_i$ is a weight associated with each background object.
We encounter here a situation similar to the cosmic
velocity field statistics (see discussion in Bernardeau \& van de
Weygaert 1996). To get a weight inversely  proportional to the local
density, there are a priori different possible schemes. One can for
instance use a two step filtering, with a usual filtering scheme on
a small grid, and then a  final filtering on the larger scale,
corresponding to the actual smoothing scale of the statistical
analysis. One could also think of using similar approaches as the
Voronoi and Delaunay methods developed by Bernardeau \& van de
Weygaert. In such a case the weigth $w_i$ would be
for instance the inverse of the surface of the Voronoi cell
occupied by a given galaxy. It would lead to a proper 'surface-average'
filtering scheme,
\beq{
\kappa^{\rm surf.}_{\theta_0}=-{3\over 2}\,\int\d^2\gam\,W_{\theta_0}(\gam)\,
{\int_{0}^{2}\d \mD_s\,\int_{0}^{\mD_{s}}
\d \mD\,{\mD\,(\mD_s-\mD)\over a\,\mD_s}\,
\delta_{\rm mass.}(\mD,\gam)\,n_s(\mD_s)\,(1+\delta_s(\mD_s,\gam))\over
\int_0^2\d\mD_s\,n_s(\mD_s)\left[1+\delta_s(\mD_s,\gam)\right]}.
\label{surf2}
}
Unlike in the velocity case, the dependence on the
local tracer density contrast does
not vanish, because of the finite width of the redshift distribution
of the sources.   In all these formulae both the density contrast of
the cosmic density field and of the sources are present. They are both
random fields and moreover their cross-correlations are a priori
comparable to their autocorrelation properties. The aim of
this study is to investigate the consequence of such couplings.

\subsection{The physical effects induced by source clustering}

First of all let me stress that the effects discussed
here are different in nature from other couplings that
have been described previously in the literature. 
A possible source a coupling between the lens and the source 
population is, for instance
the amplification effect that is expected to affect the local number
density of detected tracers (Broadhurst et al. 1995). 
This effect is due
to the fact that the apparent size of background objects depends
on the local gravitational amplification
making them more, or less, easily detectable. This effect 
is always present, even if the sources are not clustered. It is
also expected to vanish when the sources are selected on
surface brightness criteria only. Here I assume that the
sources are, in this sense, perfectly selected, so that the amplification
effect is entirely negligible. 

What are then the effects of source clustering?  It comes from the
fact that the source 'plane' is actually thick and inhomogeneous. One
mechanism originates from the fact that there might be a significant
overlapping between the  distribution of sources and the distribution
of lenses.  Let me imagine now for instance there is a large potential
well at a very large distance, in the overlapping area. Then because
the source galaxies trace somehow the matter field, you expect to have
{\it at the same time} a relative {\it excess} of close sources.  The
presence of these sources tends to reduce the gravitational  signal of
the remote lenses. One expects then that the  gravitational
distortions of the furthest lenses will be  systematically
underestimated.   In Fig. 1 a sketch of the actual situation is
proposed.

\begin{figure}
\psfig{file=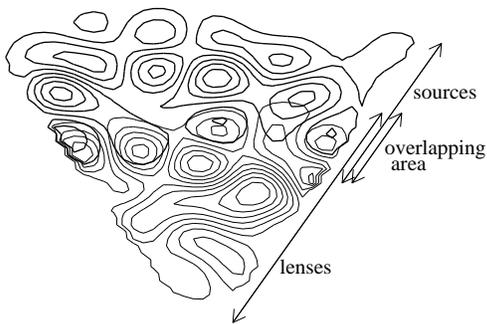, width=8 cm}
\hfill      \parbox[b]{9cm}{
\caption{Sketch of the 
lens (thin lines) and source (thick lines) density fluctuations. 
The source distribution is not expected to be smooth. Moreover the
source density fluctuations are correlated to the lens density
fluctuations. This is particularly important when the overlapping
area is large.
}}
\end{figure}

The implications of such a coupling depends on the level of description
one wishes for the local convergence. At linear order the expression
of the local convergence is not affected. Therefore the variance is
not expected to be much changed at large scale.
At small scales, the extra couplings with the source density
fluctuations can compete with the intrinsic non-linear
evolution of the projected density. It should then be addressed 
with a complete numerical study.

Another possible effect is due to the fact that the source
plane is expected to be 'bumpy', i.e. the average distance
of the sources may vary from one direction to another. For instance
if one observes the gravitational distortion induced
by a perfectly round potential with a 'bumpy' source plane,
the efficiency of the gravitational effect is expected to 
vary from one direction to another creating apparent 
substructures in the potential well.
The importance of this effect once again depends
on the level of description ones whishes. It is expected
in general to create more power at small scale, but it can be
significant (if it is present at all) only at small angular scale.
Perturbatively, this effect is expected to play 
a role only for the kurtosis and moments of higher order.
Note that contrary to the previous mechanism, this effect
is not due to the cross-correlation between the 
lenses and the sources, but to the intrinsic correlation properties
of the sources.

\subsection{Models for lens and source correlations}

The aim of the coming sections is to investigate
the implications of source clustering on high-order moments 
by means of Perturbation Theory.
Calculations can be pursued only with a 
model for the mass-galaxy and galaxy-galaxy correlation functions.
In the following I will assume that a local bias holds
and that the local galaxy density contrast can be expanded in 
terms of the {\it linear} mass density contrast,
\beq{
\delta_{\rm s}(\mD,\gam)=
b_1(\mD)\,\delta^{(1)}_{\rm mass}(\mD,\gam)+\,
b_2(\mD)\,\left[\delta^{(1)}_{\rm mass}(\mD,\gam)\right]^2+\dots
}
Although there is no complete justification for such an assumption,
it is a quite natural from the linear
analysis of Bardeen et al. (1986) or even  from the non-linear 
description\footnote{It implies in particular that the 3-point correlation 
function of the galaxy field can be exactly factorized in products
of 2-point functions} proposed by Bernardeau \& Schaeffer (1992).
Note that a skewness of about 3 (Bouchet et al. 1993, Kim \& Strauss 1997)
for the galaxies implies, $b_2\approx 0.5\,b_1^2$.
Furthermore I have to assume a given function  for the evolution
of the bias factors $b_1$ and $b_2$ with redshift. In the following I will
assume that 
\beq{
b_1(\mD)\propto 1/a(\mD)\ \ \ {\rm and}\ \ \ b_2(\mD)\propto 1/a^2(\mD).
}
According to the results of Bardeen et al. (1986) it corresponds to 
objects defined with a fixed threshold in units of the variance.

Finally the mass (and galaxy) density field is fruitfully
described by its Fourier transform,
\beq{
\delta(\mD,\gam)=\int{\d^3\vk\over (2\pi)^{3/2}}\,
\delta(\mD,\vk)\,\exp[\ii(\mD\vk_{\ortho}\cdot\gam+\mD\,k_r )]
}
where $k_r$ is the radial part of the wave vector $\vk$
and $\vk_{\ortho}$ is its perpendicular part.
In the linear regime, the Fourier components $\delta(\mD,\vk)$
grow like $a(\mD)$ for an Einstein-de Sitter Universe, and they
are assumed to obey a Gaussian statistics characterized by the
power spectrum $P(k)$,
\beq{
\delta^{(1)}(\mD,\vk)=a(\mD)\,\delta_{\rm init.}(\vk);\ \ \ 
\mg\delta_{\rm init.}(\vk)\,\delta_{\rm init.}(\vk')\md=
\delta_{\rm Dirac}(\vk+\vk')\,P(k).
}
As the numerical calculations will be done at a fixed smoothing
scale, it is reasonable to assume that $P(k)$ follows
a power law behavior,
\beq{
P(k)\propto k^n,
}
and at the scales of interest we expect (see Bernardeau et al. 1997),
\beq{
n\approx -1.5.
}

As mentionned before, from a Perturbation Theory point of view,
the variance is not affected at leading order by the source clustering.
In the following, the calculation will concentrate on the skewness
and kurtosis of the local convergence.

\section{Implications for the skewness}

At large scale, for Gaussian initial conditions, the third moment 
is given by a combination of the first order and second order
terms of the local convergence with respect to the initial 
density field (e.g. Peebles 1980, Fry 1984, Goroff et al. 1986),
\beq{
\mg\kappa^3\md\approx3\,\mg\left[\kappa^{(1)}\right]^2\,\kappa^{(2)}\md.
}
In the following, it is assumed that the effect of multiple lenses
and spurious observational couplings are negligible (see Bernardeau et al.
1997).

\subsection{Expressions of the first and second order convergence}

The presence of source clustering does not change the
expression of the first order term,
\beq{
\kappa^{(1)}_{\theta_0}=-{3\over 2}\,\int\d^2\gam\,W_{\theta_0}(\gam)\,
\int_{0}^{2}\d \mD_s\,
\int_{0}^{\mD_{s}}
\d \mD\,{\mD\,(\mD_s-\mD)\over \mD_s}\,
{\delta^{(1)}_{\rm mass}(\mD,\gam)\over a(\mD)}\,n_s(\mD_s).
}
Written in terms of $\delta(\vk)$ it reads,
\beq{
\kappa^{(1)}_{\theta_0}=-{3\over 2}\,\int_{0}^{2}\d \mD_s\,
\int_{0}^{\mD_{s}}
\d \mD\,{\mD\,(\mD_s-\mD)\over \mD_s}\,
n_s(\mD_s)\int{\d^3\vk\over (2\pi)^{3/2}}
\,\delta_{\rm init.}(\vk)\,\exp[\ii\,k_r\,\mD]\,
W(\vk_{\ortho}\,\mD\theta_0),
}
where the filter $W$ is expressed here in Fourier space.
One can further simplify this expression by introducing the
efficiency function, $\omega(\mD)$, defined by
\beq{
\omega(\mD)={3\over 2}
\int_{\mD}^2\d\mD_s\,{(\mD_s-\mD)\mD\over \mD_s}\,n_s(\mD_s),
}
so that
\beq{
\kappa^{(1)}_{\theta_0}=-\int_{0}^{2}
\d \mD\, \omega(\mD)\int{\d^3\vk\over (2\pi)^{3/2}}
\,\delta_{\rm init.}(\vk)\,\exp[\ii\,k_r\,\mD]\,
W(\vk_{\ortho}\,\mD\theta_0).
}

When the source clustering is neglected,
the second order term of the local convergence is given by
\bea{
\kappa^{(2)}_{\theta_0}=-\int_{0}^{2}
\d \mD\, \omega(\mD)
\int{\d^3\vk\,\d^3\vk'\over (2\pi)^{3}}\,F_2(\vk,\vk')
\,{a(\mD)\,\delta_{\rm init.}(\vk)\delta_{\rm init.}(\vk')}
\exp[\ii\,(k_r+k'_r)\,\mD]\,
W[\vert\vk_{\ortho}+\vk'_{\ortho}\vert\,\mD\theta_0],
}
where $F_2$ is an homogeneous function of the wave vectors
(e.g. Goroff et al. 1986).

When source clustering is taken into account, two extra terms for the
second order should be added,
\bea{
&&\kappa^{\rm s.c. (2)}_{\theta_0}=
\kappa^{(2)}_{\theta_0}-\nonumber\\
&&{3\,b_1\over 2}\int_0^2\d\mD_s\int_0^{\mD_s}\d\mD\,
n_s(\mD_s)\,{(\mD_s-\mD)\mD\over\mD_s}\,
\int{\d^3\vk\,\d^3\vk'\over (2\pi)^{3}}\,
\delta(\vk)\,\delta(\vk')\,\exp[\ii\,(k_r\mD+k'_r\mD_s)]\,
W[\vert\mD\vk_{\ortho}+\mD_s\vk'_{\ortho}\vert\,\theta_0]+\nonumber\\
&&b_1\int_0^2\d\mD\,\omega(\mD)\int_0^2\d\mD_s
n_s(\mD_s)\,
\int{\d^3\vk\,\d^3\vk'\over (2\pi)^{3}}\,
\delta(\vk)\,\delta(\vk')\,\exp[\ii\,(k_r\mD+k'_r\mD_s)]\,
W[\vert\mD\vk_{\ortho}+\mD_s\vk'_{\ortho}\vert\,\theta_0],
}

In the following, the moments will be calculated with an 
angular top-hat filter, that in $k$-space reads, 
\beq{
W(k)=2{J_1(k)\over k},
}
where $J_1$ is the spherical Bessel function.

\subsection{Expression of the variance}

The variance can be calculated straightforwardly from the
expression of the linear convergence. Using the small angle approximation
we have (Bernardeau et al. 1997),
\beq{
\mg\kappa^2\md={\Gamma[(1-n)/2]\,\Gamma[1+n/2]\over 
\Gamma[1-n/2]\,\Gamma[2-n/2]\,\pi^{3/2}}\ \theta_0^{-(n+2)}\int_0^2\d \mD\,
\omega^2(\mD)\,\mD^{-(n+2)}\equiv {\Gamma[(1-n)/2]\,\Gamma[1+n/2]\over 
\Gamma[1-n/2]\,\Gamma[2-n/2]\,\pi^{3/2}}\ \theta_0^{-(n+2)} \,I_2.
}

\subsection{Expression of the skewness}

For a top-hat filter, using the small angle approximation
and when the source clustering is neglected, the skewness
is given by (Bernardeau et al. 1997),
\bea{
s_3&\equiv&{\mg\kappa^3\md\over\mg\kappa^2\md^2}\\
s_3&=&-\left[{{36}\over 7} - {{3\,(n+2)}\over 2}\right]\ 
{\int_0^2\d\mD\,\omega^3(\mD)\,\mD^{-2(n+2)}\,a(\mD)/ I_2^2}.
}
This result has been obtained from specific properties
of the angular top-hat filter (see Bernardeau 1995). For this filter
no further approximation than the small angle approximation
is required. To compute the
skewness taking into account the source clustering, it is
of interest to assume that,
\beq{
{1\over 2\pi}
\int_0^{2\pi}\sin(\theta)\d\theta\,W\left(\vert\vk+\vk'\vert\right)=
W(k)\,W(k')
}
where $\theta$ is the angle between the wave vectors $\vk$
and $\vk'$. This property is not exact, but the error it induces
is extremely weak, less than 1\% for $n\approx -1.5$.
This property implies in particular that the two filtering
schemes Eqs. (\ref{surf1}) and (\ref{surf2}) give the same results 
for top-hat filtering.

Taking advantage of this expression, we have,
\bea{
s_3^{\rm s.c.}=s_3-&&{9\over I_2^2}\,b_1\,
\int_0^2\d\mD\int_0^2\d\mD'\,
\omega(\mD)\,{\mD(\mD'-\mD)\over\mD'}\,n_s(\mD')\,\omega(\mD')\,
\mD^{-(n+2)}\,\mD'^{-(n+2)}+\nonumber\\
&&{6\over I_2}\,b_1\,{\,\int_0^2\d\mD\,
\omega(\mD)\,n_s(\mD)\,\mD^{-(n+2)}/ I_2},
}
when source clustring is taken into account.
It of course depends only on $b_1$.
In Table 1, I give the results for the source models described in Sect. 5.

\section{The kurtosis}

In the frame of Perturbation Theory, the expression of the kurtosis 
is given by
\bea{
\mg\kappa^4\md_c\equiv\mg\kappa^4\md-3\,\mg\kappa^2\md^2=
6\,\mg\left[\kappa^{(1)}\right]^2\left[\kappa^{(2)}\right]^2\md_c+4\,
\mg\left[\kappa^{(1)}\right]^3\kappa^{(3)}\md_c.
}
When the clustering effects is neglected the third order $\kappa$
is given by the integral over the line-of-sight of the local 
third order density. 
Then, using the general formulae of Bernardeau (1995) one can compute
the kurtosis,
\bea{
s_4&\equiv&{\mg\kappa^4\md_c\over
\mg\kappa^2\md^3};\\
s_4&=&
\left[{{2540}\over {49}}-33\,(n+2)+{{21\,{(n+2)^2}}\over4}\right]
{\int_0^2\d\mD\,\omega^4(\mD)\,\mD^{-3(n+2)}\,a^2(\mD)/ I_2^3}.
}

The general expression of the third order moment contains extra terms,
\bea{
\kappa^{\rm s.c. (3)}=&&
\kappa^{(3)}-{3\over 2}\int_0^2\d\mD_s\int_0^{\mD_s}\d\mD\,
{(\mD_s-\mD)\mD\over\mD_s}\,\delta^{(2)}_{\rm mass.}(\mD)\,
n_s(\mD_s)\,\delta^{(1)}_{\rm s}(\mD_s)-\nonumber\\
&&{3\over 2}\int_0^2\d\mD_s\int_0^{\mD_s}\d\mD\,
{(\mD_s-\mD)\mD\over\mD_s}\,\delta^{(1)}_{\rm mass.}(\mD)\,
n_s(\mD_s)\,\delta^{(2)}_{\rm s}(\mD_s)+\nonumber\\
&&\int_0^{2}\d\mD\,\omega(\mD)
\delta^{(2)}_{\rm mass.}(\mD)\,
\int_0^2\mD_s\,n_s(\mD_s)\,\delta^{(1)}_{\rm s}(\mD_s)+\nonumber\\
&&{3\over 2}\int_0^2\d\mD_s\int_0^{\mD_s}\d\mD\,
{(\mD_s-\mD)\mD\over\mD_s}\,\delta^{(1)}_{\rm mass.}(\mD)\,
\delta^{(1)}_{\rm s.}(\mD_s)\,
n_s(\mD_s)\,\int_0^2\mD_s\,n_s(\mD_s)\,\delta^{(1)}_{\rm s}(\mD_s)
+\nonumber\\
&&\int_0^{2}\d\mD\,
\omega(\mD)\,\delta^{(1)}_{\rm mass.}(\mD)\,
\left(
\int_0^2\mD_s\,n_s(\mD_s)\,\delta^{(2)}_{\rm s}(\mD_s)-
\left[\int_0^2\mD_s\,n_s(\mD_s)\,\delta^{(1)}_{\rm s}(\mD_s)\right]^2
\right).
}

As a result, when one wants to take fully into account the source-clustering
effects, many new terms should be included.

In order to simplify the expression of the resulting kurtosis,
I introduce the function,
\bea{
\mV(\mD,\mD')&=&-{3\over 2}
\mH(\mD'-\mD')\,{(\mD'-\mD)\mD\over\mD'}\,n_s(\mD')+
\omega(\mD)\,n_s(\mD'),
}
where $\mH$ is the Heaviside function, $\mH(\mD'-\mD)=0$ if $\mD' < \mD$
and $\mH(\mD'-\mD)=1$ if $\mD'\ge\mD$.

Then, $s_4^{\rm s.c.}$ is given by
\bea{
s_4^{\rm s.c.}=s_4-
&&{4\over I_2^3}\,b_1\,\left[{36\over 7}-{3\over2}(n+2)\right]\,
\int_0^2\d\mD_1\,\omega(\mD_1)\,\mD_1^{-(n+2)}
\int_0^2\d\mD_2\,\omega^2(\mD_2)\,
\left[\mV(\mD_1,\mD_2)+\mV(\mD_2,\mD_1)\right]
\mD_2^{-2(n+2)}+\nonumber\\
&&
{12\over I_2^3}\,b_1^2\,
\int_0^2\d\mD_1\,\omega(\mD_1)\,\mD_1^{-(n+2)}
\int_0^2\d\mD_2\,
\left[\mV(\mD_1,\mD_2)+\mV(\mD_2,\mD_1)\right]\,
\mD_2^{-(n+2)}\times\nonumber\\
&&\ \ \ \ \ \ \ \int_0^2\d\mD_3\,
\left[\mV(\mD_2,\mD_3)+\mV(\mD_2,\mD_3)\right]\,
\mD_3^{-(n+2)}\,\omega(\mD_3)-\nonumber\\
&&
{24\over I_2^3}\,b_2\,
\int_0^2\d\mD_1\,\omega(\mD_1)\,\mD_1^{-(n+2)}
\int_0^2\d\mD_2\,\omega^2(\mD_2)\,\mV(\mD_1,\mD_2)\,
\mD_2^{-2(n+2)}-\nonumber\\
&&{24\over I_2^3}\,b_1^2\,
\int_0^2\d\mD_1\,\omega(\mD_1)\,\mD_1^{-(n+2)}
\int_0^2\d\mD_2\,\omega(\mD_2)\,n_s(\mD_2)\,\mD_2^{-(n+2)}\times\nonumber\\
&&\ \ \ \ \ \ 
\int_{\mD_1}^2\d\mD_3\,\omega(\mD_3)\,n_s(\mD_3)\,\mD_3^{-(n+2)}\,
\left[{3\over 2}\,{(\mD_3-\mD_1)\mD_1\over\mD_3}-\omega(\mD_1)\right]
}

Note that, contrary to the skewness case, there is a contributing term
\beq{
 s_4^{\rm bumps}\equiv{12\over I_2^3}\,b_1^2\,
\int_0^2\d\mD_1\,\omega(\mD_1)\,\mD_1^{-(n+2)}
\int_0^2\d\mD_2\,\mV(\mD_1,\mD_2)\,
\mD_2^{-(n+2)}\int_0^2\d\mD_3\,\mV(\mD_2,\mD_3)\,\mD_3^{-(n+2)}\,
\omega(\mD_3), 
} 
which is due to the source auto-correlation
function only.  It corresponds to the second mechanism described in
Sect. 2.2.  This term does not disappear a priori when the
overlapping between the source distribution and the lens efficiency
function is arbitrarily small. The results shown in table 1 
prove however that in practice this contribution is always
negligible compared to the cross-correlation effects.

\section{Discussions}

\subsection{Numerical results}

\begin{figure}
\psfig{file=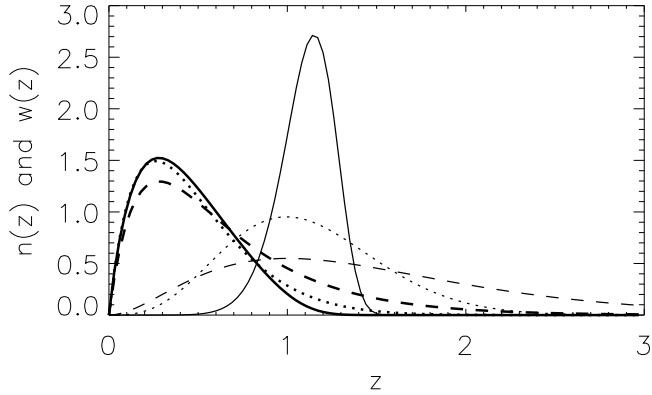, width=10 cm}
\hfill      \parbox[b]{7cm}{
\caption{Shapes of the
distribution functions of the sources (thin lines) and 
lens efficiency functions (thick lines) for the three models 
(models 1, 2, 3 are respectively plotted with solid, 
dotted and dashed lines). All functions are arbitrarily normalized to
unity
}}
\end{figure}

In order to get numerical results one should choose a specific 
model for the redshift source distribution. I will
assume that $n_s$ takes the form,
\beq{
n_s(z)\propto z^{\alpha}\,\exp\left[-(z/z_0)^\beta\right].
}
I make the calculations for three hypothesis,
\bea{
{\rm model\ 1:\ }&&z_0=1.15,\ \ \ \alpha=8.,\ \ \ \beta=8.;\\
{\rm model\ 2:\ }&&z_0=0.75,\ \ \ \alpha=3.,\ \ \ \beta=1.8;\\
{\rm model\ 3:\ }&&z_0=0.5,\ \ \ \alpha=2.,\ \ \ \beta=1.
}
For the first two models the mean redshift of the sources is about unity,
with a larger distribution in the second case. The first model
has somewhat arbitrary parameters, whereas the second is motivated by
results of galaxy evolution models (Charlot \& Fall, in preparation). 
It would correspond to galaxies
with a $I$ magnitude between 22 and 24. It has a redshift 
dispersion higher than model 1.
In model 3, I have assumed a very broad distribution of the redshift 
distribution. 
The resulting values for the skewness and the kurtosis in such
models are given in table 1. As expected the first two models
give roughly the same answers, because the mean source
redshifts are very close. In the third model there is
a significant number of sources at very high redshift. It lowers
the values of $s_3$ and $s_4$.

\begin{table}
\caption{Skewness and kurtosis with source clustering effects
}
$
\begin{array}{llll}
\hline
\vspace{.1cm}
                  & {\rm model\ 1}  & {\rm model\ 2}  & {\rm model\ 3}   \\
\mg z_s\md\ \     &1.11             &1.11             &1.50\\
\vspace{.1cm}
\Delta(z_s)\ \    &0.15             &0.42             &0.87\\
s_3^{\rm s.c.}\ \ & -37.6+1.05\,b_1 &-39.9+6.4\,b_1  & -29.1+7.8\,b_1  \\
s_4^{\rm s.c.}\ \ & 2890-  49\,b_1+ 12\,b_1^2-  24\,b_2\ \ \  &
                     3330- 410\,b_1- 217\,b_1^2- 280\,b_2\ \ \ & 
                     1790- 390\,b_1-  4\,b_1^2- 290\,b_2\\
\vspace{.1cm}
s_4^{\rm bumps}\ \ & 1.3\,b_1^2\ \ \  & 19\,b_1^2 \ \ & 21\,b_1^2\\
\hline
\end{array}
$
\end{table}

Skewness and kurtosis show a similar sensitivity with
source clustering effects. For the skewness, the remaining corrective 
term can be  as large as 25\% of the signal when the 
redshift distribution of the sources is large. The correction
is about 30\% for the kurtosis assuming that $b_1\approx 1$ and
$b_2 \approx 0.5$. However for a narrow redshift distribution, the 
corrective terms remain small (about 3\% for model 1 for the skewness and 2\%
for the kurtosis). It shows that the high order moments are slightly 
dependent on source clustering but that this dependence can be controlled 
if the redshift dispersion of the sources is low enough.
If one does not want to include pre-knowledge on the
galaxy-mass cross-correlation to analyze the data, it will 
be important to reduce as much as possible the width of the
redshift distribution in the adopted selection criteria.
Note that it will be anyway possible to get constraints on the
mass-source and source-source correlations from counts in cells 
statistics applied to the selected sources (Schneider 1997,
van Waerbeke 1997).

It is also worth to have in mind that these results 
depend on the shape of the filter. In particular if
one does not use a top-hat filter the two filtering schemes
are not equivalent. This would be the case in particular 
for compensated filters.  The ``proper'' surface-weighting scheme
is expected to be, in general, less sensitive to the source clustering. 

\subsection{Skewness and kurtosis to measure the non-Gaussian effects}

In the preliminary investigations of the non-Gaussian
properties expected to be observable
in convergence maps, the focus has been mainly
put on the skewness. It is indeed the first non-trivial cumulant
expected to emerge due of mode couplings
and its calculation is possible in the frame of Perturbation Theory.
However, it would be stupid to limit the search of non-Gaussian
effects to the skewness only. 
Even at the level of the shape of the local convergence PDF,
the skewness cannot entirely characterize the departure from a
gaussian distribution. This departure manifests itself by 
the apparition of a whole set of cumulants. In the context
we are interested in, the PDF of $\kappa$
is essentially a 2 parameter familly (when
a given population of sources is considered): it depends on the amplitude
of the density fluctuations and on $\Omega$ (neglecting at
first view the $\Lambda$ dependence). It implies that all 
cumulants are somehow connected together. In particular the kurtosis $s_4$
appears simply to be an other way of constraining $\Omega$ (it
is in perturbation theory independent of $\sigma_8$) which means
that $s_3$ and $s_4$ are naturally related. This relation can be
easily identified from the results of Table 1. Indeed one can
check that, in the absence of source clustering, 
\beq{
{s_4\over s_3^2} \approx 2.05\ \ {\rm to}\ \ 2.1,
}
for the three models. The dependence of the
ratio $s_4/s_3^2$ on $n$ is expected to be weak, as well as its
dependence on the  cosmological parameters. 
We encounter here the same situation that has
been noticed by Bernardeau (1994) for the cumulants of the
cosmic density or the cosmic velocity divergence filtered with a 3D top-hat
window function. 
This property is a priori valid in the quasilinear regime only.
In the strongly non-linear regime the
validity of this relationship remains an open question
although results obtained in a phenomenological description
of the cumulant behavior of the 3D cumulants (Colombi et al. 1997)
strongly suggests that such a relation may survive in the
non-linear regime. The consequences are two fold,
\begin{itemize}
\item $s_4$ might be used as an alternative method
of measuring $\Omega$; 
\item the fact that $s_4/s_3^2$ should be
approximatly 2 in all cases might also reveal a precious property to use 
for testing the correctness of observational results. For instance, when
the bias effects are too strong this ratio tends indeed to increase up to 
about 2.7. 
\end{itemize}

\subsection{Systematics and spurious couplings}

To summarize, the possible spurious couplings that may affect
the statistical properties of the convergence maps 
that have been identified so far are:
\begin{itemize}
\item Multiple lens couplings. This is due to the fact that
the convergence effects of lenses that are aligned 
do not add linearly. Actually, to compute the effect of
combined lenses one should multiply the amplification
matrices. Departure from the Born approximation appears at the same
level of approximation. Both are shown to induce at most a few percent
effect correction on the skewness (see Bernardeau et al. 1997).
\item Magnification effects. This is due to the fact that the
population of selected sources may depend
on the local magnification (see again Bernardeau et al. 1997). This
effect is difficult to quantify a priori. It should be tested with the
selection algorithms which are actually used.
\item Source density fluctuations. This is one of the effect
which is investigated in this paper. It comes from the fact that
the source plane appears ``bumpy''. At large scale, it induces
corrective terms at the level of the kurtosis only, but it might 
play a more important role at small scale.
\item Source-lens correlations. This effect, the main effect 
investigated in this paper, is due to the fact that the sources
might also somehow trace the foreground lenses when the distribution
function of the source redshift is large enough. The latter two effects are
shown to be negligible when the width of the redshift distribution is small
enough (say less than 0.15). 
\end{itemize}
All these effects have been shown to intervene for the
high order moments only, in the frame
of perturbation theory that is at large enough scale
(above about 10 arcmin scale).
In this approach it is
possible to estimate the magnitude of these effects and to show that 
for a reasonable redshift distribution they
are negligible. 

In practice however it is likely
that the scales of interest will be much smaller than scale
at which the perturbation theory regime is valid. It would therefore
be interesting to extend these results to the small
scales. In particular the effect of source clustering on the variance
might not be totally negligible.

\section*{Acknowledgments}
The author would like to thank Yannick Mellier and
Ludovic van Waerbeke for many discussions
and the referee, Bhuvnesh Jain, for
suggestions that have contributed to the
improvement of the manuscript. The author is also 
grateful to IAP where most of this work has been conducted.

\end{document}